\documentclass[twocolumn,nopacs,floatfix,superscriptaddress]{revtex4}
\usepackage{amsmath}
\usepackage{graphicx}
\usepackage{epsfig}
\usepackage{color}
\usepackage{soul}
\usepackage{longtable}
\usepackage{verbatim}
\usepackage[OT1]{fontenc}
\usepackage[normalem]{ulem}

\begin{document}
\title{Scale free density and correlations fluctuations in the dynamics of large microbial ecosystems}

\author{Nahuel Zamponi}
\email{zamponi.n@gmail.com}
\affiliation{Division of Hematology and Medical Oncology, Department of Medicine, Weill Cornell Medicine. New York, NY, USA}
\author{Tomas S. Grigera}
\affiliation{Instituto de F\'isica de L\'iquidos y Sistemas Biol\'ogicos (IFLYSIB), CONICET y Universidad Nacional de La Plata, La Plata, Argentina}
\affiliation{Departamento de F\'isica, Facultad de Ciencias Exactas, Universidad Nacional de La Plata, La Plata, Argentina}
\affiliation{Istituto dei Sistemi Complessi, Consiglio Nazionale delle Ricerche, via dei Taurini 19, 00185 Rome, Italy}
 \affiliation{Consejo Nacional de Investigaciones Cient\'{\i}fcas y Tecnol\'ogicas (CONICET), Buenos Aires, Argentina}
 
 \author{Ewa Gudowska-Nowak}
\affiliation{Mark Kac Center for Complex Systems Research and Institute for Theoretical Physics, Jagiellonian University, Krak\'ow, Poland}

\author{Dante R. Chialvo}
\affiliation{Instituto de Ciencias F\'isicas (ICIFI-CONICET), Center for Complex Systems and Brain Sciences (CEMSC3), Escuela de Ciencia y Tecnolog\'ia, Universidad Nacional de Gral. San Mart\'in, Campus Miguelete, San Mart\'in, Buenos Aires, Argentina}
\affiliation{Consejo Nacional de Investigaciones Cient\'{\i}fcas y Tecnol\'ogicas (CONICET), Buenos Aires, Argentina}
\affiliation{Mark Kac Center for Complex Systems Research and Institute for Theoretical Physics, Jagiellonian University, Krak\'ow, Poland}

 \date{\today}

\begin{abstract}
Microorganisms self-organize in very large communities exhibiting complex fluctuations. Despite recent advances, still the mechanism by which these systems are able to exhibit large variability at the one hand and dynamical robustness on the other, is not fully explained. With that motivation, here we analyze three aspects of the dynamics of the microbiota and plankton: the density fluctuations, the correlation structure and the avalanching dynamics. In all communities under study we find that the results exhibits scale-free density fluctuations, anomalous variance' scaling, scale-free abundance correlations and stationary scale-free avalanching dynamics. These behaviors, typical in systems exhibiting critical dynamics, suggest criticality as a potential mechanism to explain both the robustness and (paradoxical) high irregularity of processes observed in very large microbial communities.
\end{abstract}

\maketitle

\section{Introduction}

Microbes are ubiquitous in nature and essential for ecosystems and human health. Despite efforts to characterize the structure of these communities, the mechanisms that determine their complex dynamics remain unknown~\cite{costello, cho, koskella}. In this paper we present evidence for critical dynamics of microbial communities from distinct environments. Our findings suggest that criticality may explain the scale-free fluctuations and power law tail distributions observed in the data.

Here we focus on two different microbe communities: human microbiota~\cite{caporaso, david} and plankton~\cite{martin-platero}. Recent work has explored the statistical properties of these systems focusing on the description of macroecological laws, accounting for the structure of these ensembles which may explain the inter- and intra-host dynamics~\cite{servan, grilli, ji, dalziel, zaoli, shoemaker, wang}. The goal of the present work is different: we attempt to demonstrate that the microbiome and plankton are microbial systems poised near criticality. Our hypothesis stems from the fact that the microbial data exhibits properties consistent with characteristics of critical systems both in theory and models.

We use scaling arguments to demonstrate that both - the statistical properties of the abundance fluctuations and correlations between taxonomic groups are consistent with critical dynamics. Microbial systems, devoid of any spatial structure (similar to mean-field descriptions), exhibit high susceptibility and large variability while attaining robustness by adopting certain taxonomic associations via functional interactions uncovered by our finite-size scaling correlation analysis.

The paper is organized as follows. In Sec. II we introduce the data and describe the basic statistics concerning abundance distribution and the fluctuation of temporal averages. In Sec. III we focus on the collective properties of these systems by studying the correlation matrix. To connect with a usual observations in critical phenomena, in Sec. IV we explore the scaling of avalanches of activity in both the microbiome and plankton data, as well their stationarity. We close the paper with a short discussion about the implications of our findings.
The Appendix includes a null model used to reconstruct synthetic time series of abundances, testing whether first order autocorrelation suffices to explain the statistical properties of the real data.
%%%%%%%%%%%%%%%%%%%%% FIG1 %%%%%%%%%%%%%%%%%%%%%%%%%%%%

\begin{figure*}
\centering
\includegraphics[width=.56\textwidth]{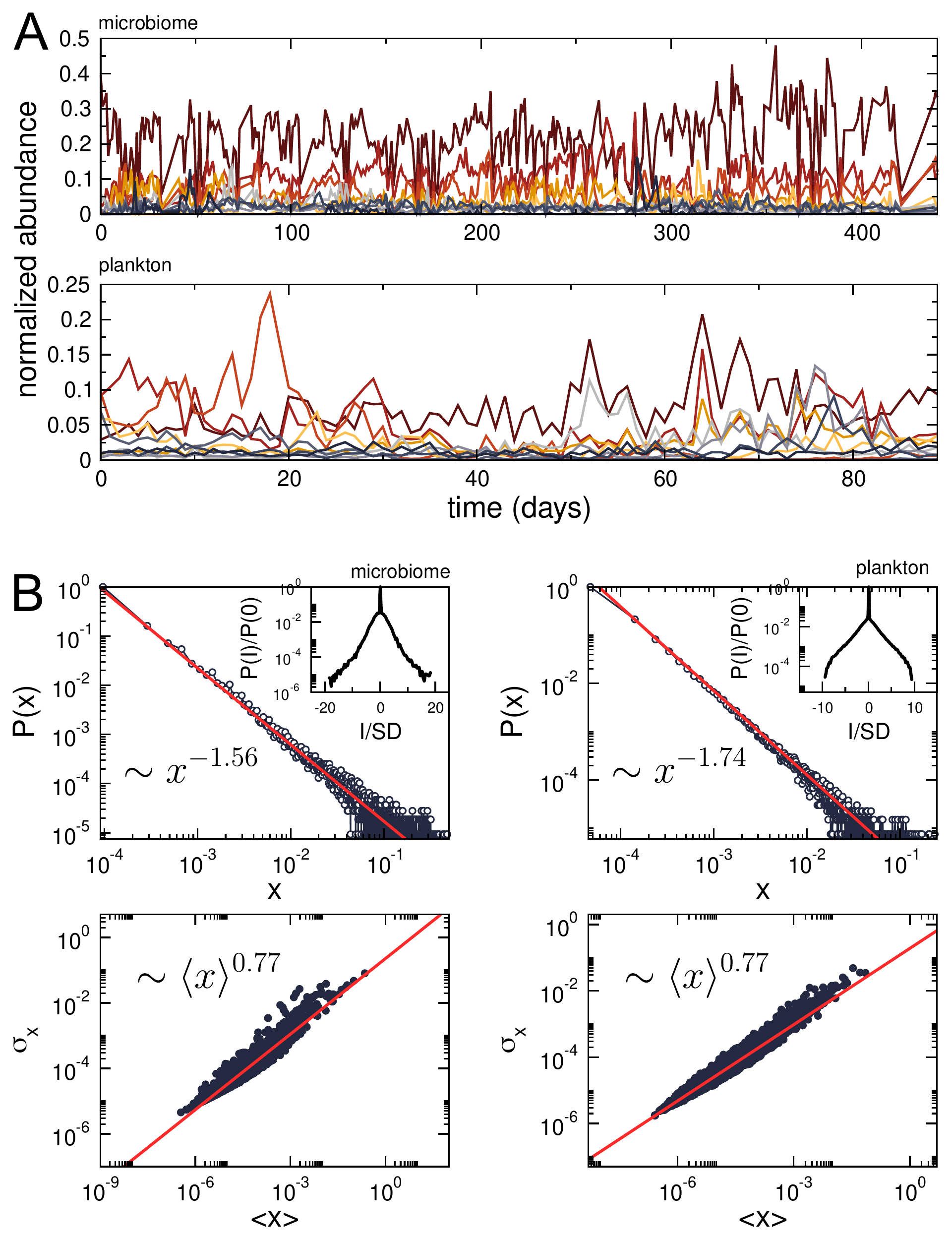}
\caption{{\bf Scale-free fluctuations of the genera' abundance in the microbiome and plankton}. A) Time series of relative abundances of bacterial genera from the microbiome (subject M3~\cite{caporaso}) and from plankton (Ref.~\cite{martin-platero}). B) Top panels: The tail distribution function of abundances across time derived for all genera decays as a power-law with exponents $-1.56$ and $-1.74$ for the microbiome and plankton, respectively. Bottom panels: The relation between mean and standard deviation for the relative abundance of bacterial genera is non-Poissonian, approximating a power-law with exponent $\alpha = 0.77$ in both cases.}
\label{fig:fig1}
\end{figure*}

\section{Density fluctuations}

We first characterized the fluctuations of the microbes abundances. Individuals are denoted by the so-called Operational Taxonomic Units (OTU), a definition used to classify groups of closely related individuals, most often grouped by DNA sequence similarity of a specific taxonomic marker gene.
Abundances show rather large fluctuations, as witnessed by the normalized abundance of the 10 most abundant bacteria genera from the microbiome and plankton as a function of time (Fig.~\ref{fig:fig1}A). As a first characterization of these fluctuations, we compute the density distribution $f(X)$ of the normalized abundances from all genera and derive the corresponding tail distribution function $Prob\{X\ge x\}\equiv\int_x^{\infty}f(X)dX=P(x)$, cf.~top panels in Fig.~\ref{fig:fig1}B: these follow power-law distributions over more than two decades, demonstrating that abundance fluctuations lack a characteristic scale. In agreement with this observation, the insets in the top panels in Fig.~\ref{fig:fig1}B show that the density distribution of time series of successive increments $I(t)=X(t+1)-X(t)$ normalized by its standard deviation $SD$ exhibit deflections from exponential tails. To corroborate the presence of nontrivial fluctuations, we computed the standard deviation of the normalized abundances across time $\sigma _{x}$ as a function of the average normalized abundance $\langle x \rangle$, a relation known as Taylor's law~\cite{eisler}. As shown in the bottom panels of Fig.~\ref{fig:fig1}B, $\sigma _{x}$ scales with $\langle x \rangle$ following a power-law with exponent $\approx 0.77$, supporting the notion that fluctuations are scale invariant and non-Poissonian in nature.
Since the presence of fluctuations with power-law distributions and anomalous scaling of its variance are common in critical phenomena~\cite{life_edge}, the conjecture can be made that its origin may be related to critical dynamics. This could reconcile the apparent contradiction observed in these systems, as being robust ecosystems despite exhibiting large variability in the abundances.

\section{Finite-size scaling of correlations}

%\subsection {Microbiome eigenvalues spectra}
 %%%%%%%%%%%%%%%% FIG2 %%%%%%%%%%%%%%%%%%%%%%%% 
\begin{figure*} 
\centering
\includegraphics[width=0.6\textwidth]{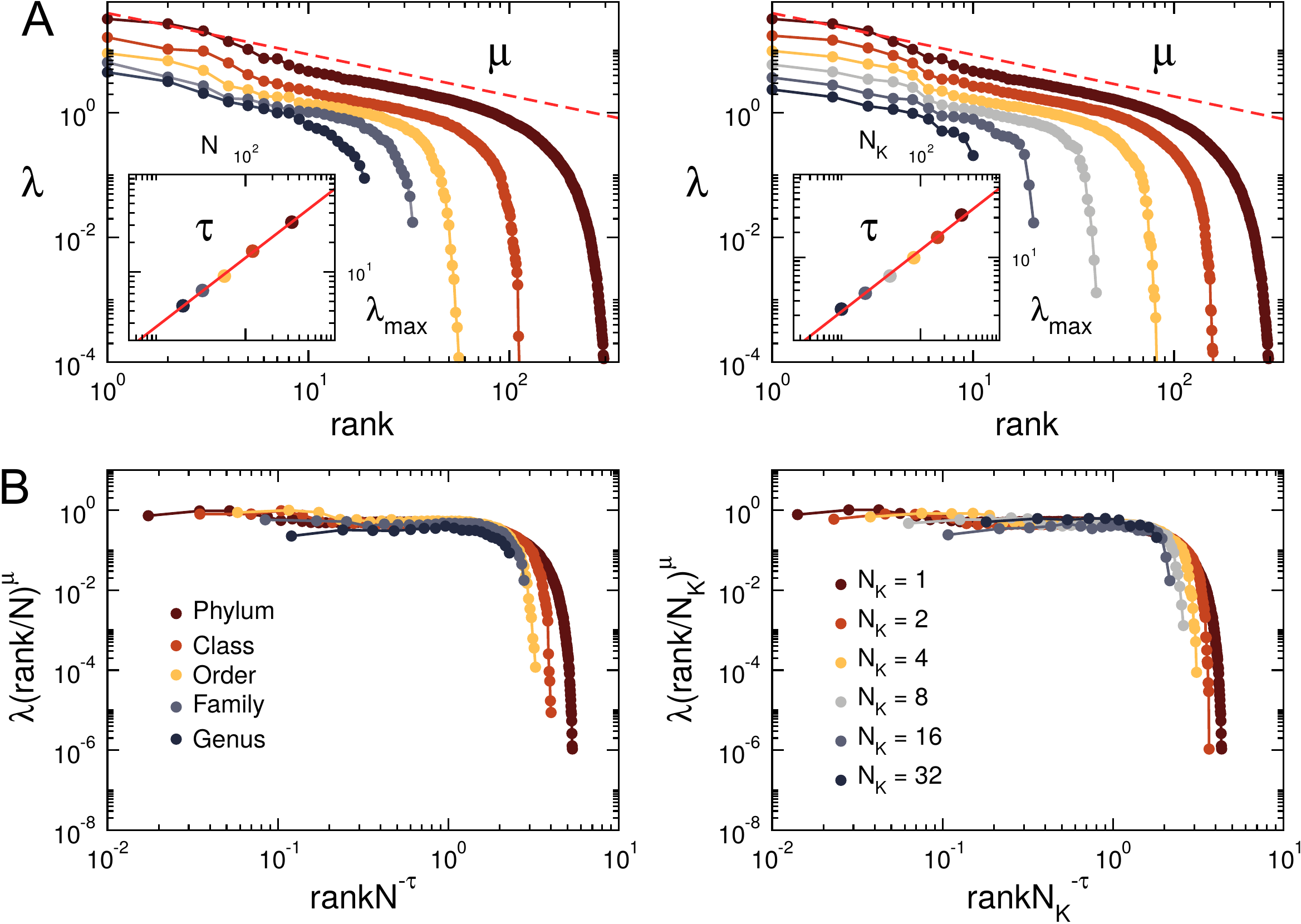}
\caption{{\bf Scaling in eigenvalues of the correlation matrix spectra (Microbiome M3).} A) power-law dependence on rank for data grouped based on taxonomy (left) or RG (right). B) Collapse of eigenvalues ranked distributions using the exponents $\mu$ and $\tau$ obtained in A)}
\label{fig:fig2}
\end{figure*}
%%%%%%%%%%%%%%%%%%%%%%%%%%%%%%%%%%%%%%%%

Next we look for signs of scaling in the correlations among the community constituents: we explore the behavior of the correlations between OTUs at different coarse graining levels. This is motivated by the expectation that correlations depend of system's size if the variability described in the previous section is originated by critical dynamics~\cite{grigera}. This analysis is restricted to the microbiome data, since the plankton time series are too short to allow us to obtain a definite correlation matrix from the available data.

We do the coarse-graining in two different ways. First, we consider the available taxonomic annotation of each OTU as follows: the smallest grain (biggest system size) was obtained by grouping OTUs by {\em Genus}, yielding a total of 329 OTUs. Intermediate grains were obtained grouping OTUs by {\em Family} (120), {\em Order} (58), and {\em Class} (33). Finally, the biggest grain (smallest system size) was obtained grouping by {\em Phylum}, yielding a total of 20 OTUs. Thus at each coarse-graining level $n$ we have a set of $N_n$ abundance time series $x_i^{(n)}(t)$, from which we obtain the correlation matrix at each level,
\begin{equation}
c_{ij}^{(n)} = \frac{\langle \delta x_i^{(n)} \delta x_j^{(n)} \rangle}{\sqrt{\langle(\delta x_i^{(n)})^2\rangle \langle(\delta x_j^{(n)})^2\rangle}},
\label{eq:corr}
\end{equation}
where $\delta x_i^{(n)} = x_i^{(n)} - \langle x^{(n)} \rangle$.

The second coarse-graining procedure is obtained using the phenomenological {\em Renormalization Group} (RG) approach recently applied to the study of neuronal correlations~\cite{meshulam}. We start with variables $\{x_i(t)\}$ describing abundances of each OTU $i = 1,2, ..., N$ at time $t$ and compute the correlation matrix, Eq.~\eqref{eq:corr}. We then search for the largest non-diagonal element of this matrix, identifying the maximally correlated pair $i,j_*(i)$, and define the coarse grained variable
\begin{equation}
	x^{(2)}_i = Z^{(2)}_i (x_i + x_{j_*(i)}),
\end{equation}
where $Z^{(2)}_i$ restores the normalization. We then remove the pair $[i, j_*(i)]$, search for the next most correlated pair, and so on, greedily, until the original $N$ variables became $[N/2]$ pairs. Iterating this process, we obtain, at coarse-graining level $k$, $N_K = [N/K]$ clusters of size $K=2^k$, represented by coarse-grained variables ${x_i^{(K)}}$.

In this way we obtain two sets of matrices, one built according previous information of taxonomic hierarchy, and another obtained from the hierarchy of correlations. We studied the ranked distribution of eigenvalues of the correlation matrices at each size (or, equivalently, coarse-graining level), since for scale-invariant systems (the case for systems near the critical point) a power-law dependence of the eigenvalues on rank is expected, as well as a scaling behavior of the eigenvalue spectra as a function of size \cite{meshulam}. As shown in Fig.~\ref{fig:fig2}A the eigenvalue vs.\ rank curves are indeed power law in both cases (with exponent $\mu \approx 0.66$ and $\approx 0.55$ for the taxonomic and RG hierarchies, respectively). Moreover, we corroborated that the magnitude of the largest eigenvalue $\lambda _{max}$ scales with the system size following a power-law with exponent $\tau$ ($\approx 0.71$ and $\approx 0.82$ for taxonomy and RG, respectively), as shown in the insets of Fig.~\ref{fig:fig2}. As we demonstrate in Fig.~\ref{fig:fig2}B, we can use these exponents ($\mu $ and $\tau$) to collapse the eigenvalue spectra in both cases, confirming the presence of scaling behavior in this system.
%%%%%%%%%%%%%%%%%%%% FIG3%%%%%%%%%%%%%%%%%%%%%%%%%%%%%%%
\begin{figure*} 
\centering
\includegraphics[width=0.85\textwidth]{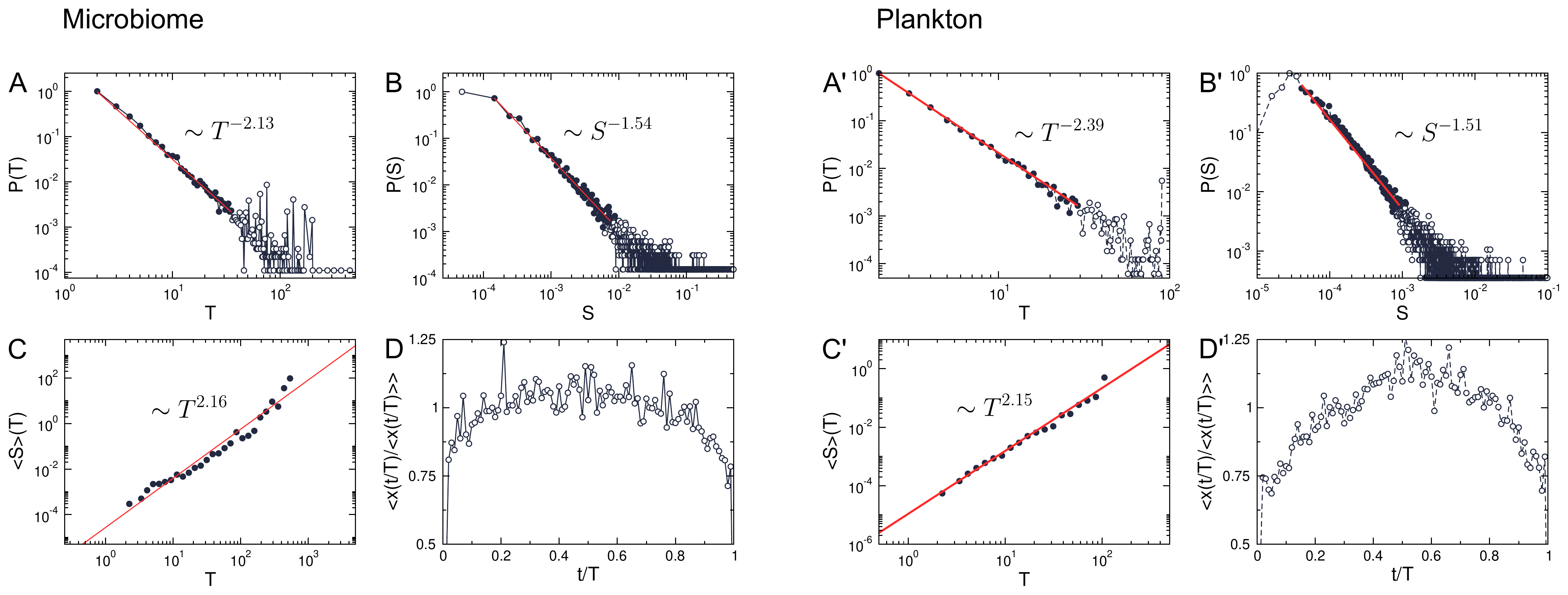}
\caption{{\bf Avalanche distributions of Microbiome M3 and Plankton Bacteria are scale free}. \emph{Left side panels: Microbiome M3}. A) The distribution of events duration (T) decays as a power-law with exponent $\alpha \approx 2.13$. B) The distribution of event sizes (S) decays as a power-law with exponent $\tau \approx 1.54$. C) The average event size as a function of event duration follows a power-law with exponent $\gamma \approx 2.16$. D) The normalized average event profile collapses to a universal function that reaches a plateau at $t/T \approx 0.5$. Filled symbols indicate the region used to fit the power-law. 
\emph{Right side panels: Plankton Bacteria}. Panels are formatted as in left-side panels. In this case we obtain $\alpha \approx 2.39$, $\tau \approx 1.51$, and $\gamma \approx 2.15$. 
}
\label{fig:fig3}
\end{figure*}
%%%%%%%%%%%%%%%%%%%%%%%%%%%%%%%%%%%%%%%%%%%%%%%%%%
It is remarkable that a biological system with no strictly defined spatial structure (i.e., suspensions in which every organism can interact with everyone else) exhibit this rich correlation structure. But more importantly, note that the correlation structure we found using the taxonomic information is identical to that produced by the hierarchy of their mutual correlations (left and right panels in Fig.~\ref{fig:fig2}B) suggesting that the evolutionary distance between OTUs may result from a symbiotic organizing principle of the microbial systems.

\section{Avalanches of microbe abundance}

Critical systems are known to exhibit bursts of activity, composed by avalanche events with peculiar statistics. We studied the statistical properties of the events duration ($T$) and size ($S$). Following previous work \cite{ji, suweis, bertuzzo, keitt} we defined an event as the time series of abundances between the emergence and subsequent disappearance of an OTU. In the following we will exemplify our analysis using the data from the microbiome. We first determined the density distribution of both event duration and size. As shown in Fig.~\ref{fig:fig3}A and B, these quantities decay as power-laws with exponents $\approx 2.13$ and $\approx 1.54$, respectively. Then, we asked how event duration and size were related and determined the average event size as a function of event duration. To estimate event size as a function of duration, event durations were binned and the average event size was then calculated on the binned durations. 

Our results shown in Fig.~\ref{fig:fig3}C indicate that event size scales with event duration following a power-law with exponent $\approx 2.16$. Besides the obvious conclusion that {\em ``the longer the event the larger its size''}, the value of the exponent suggests that the underlying mechanism generating the correlation within an event is different from a random walk (RW) and is more consistent with a branching process (BP) \cite{villegas, zapperi}. The functional relation between event duration and size shown in Fig.~\ref{fig:fig3}C implies that the time series of abundances is self-affine \cite{mandelbrot}. Therefore, we should be able to extract an average shape of the events. Fig.~\ref{fig:fig3}D shows that, after appropriate rescaling, an average shape event is extracted from the collapse of all normalized events.
%%%%%%%%%%%%%%%%%%%%%%%%%%%%%%%%%%%%%%%
Supporting the generality of these findings, a similar picture holds for plankton data (summarized in the right panels of Fig.~\ref{fig:fig3}). In summary, we have that
\begin{equation}
	P(T) \sim T^{-\alpha}, \qquad	P(S) \sim S^{-\tau}, \qquad
	\langle S \rangle (T) \sim T^\gamma,
\end{equation}
with $\alpha _{M} \approx 2.13$ and $\alpha _{P} \approx 2.39$, $\tau _{M} \approx 1.54$ and $\tau _{P} \approx 1.51$, and $\gamma _{M} \approx 2.16$ and $\gamma _{P} \approx 2.15$, where $M$ and $P$ sub-indices refer to microbiome and plankton, respectively. In a critical branching process it is expected that the following scaling relation holds among exponents,
\begin{equation}
	\frac{\alpha - 1}{\tau - 1} = \gamma.
	\label{eq:scale}
\end{equation}
Using this relation one would find a value of $\gamma$ equal to 2.09 for microbiome and 2.73 for plankton.
Considering the combined uncertainties, the values of these exponents are in good agreement with those expected for a critical system, namely, a branching process, where $\alpha = 2$, $\tau = 1.5$ and $\gamma = 2$ \cite{villegas, zapperi}.
%%%%% %%%%%%%%%%%%%%%%%%%FIG4 %%%%%%%%%%%%%%%%%%%%%%%%%%%%%% 
\begin{figure*}
\centering
\includegraphics[width=0.9\textwidth]{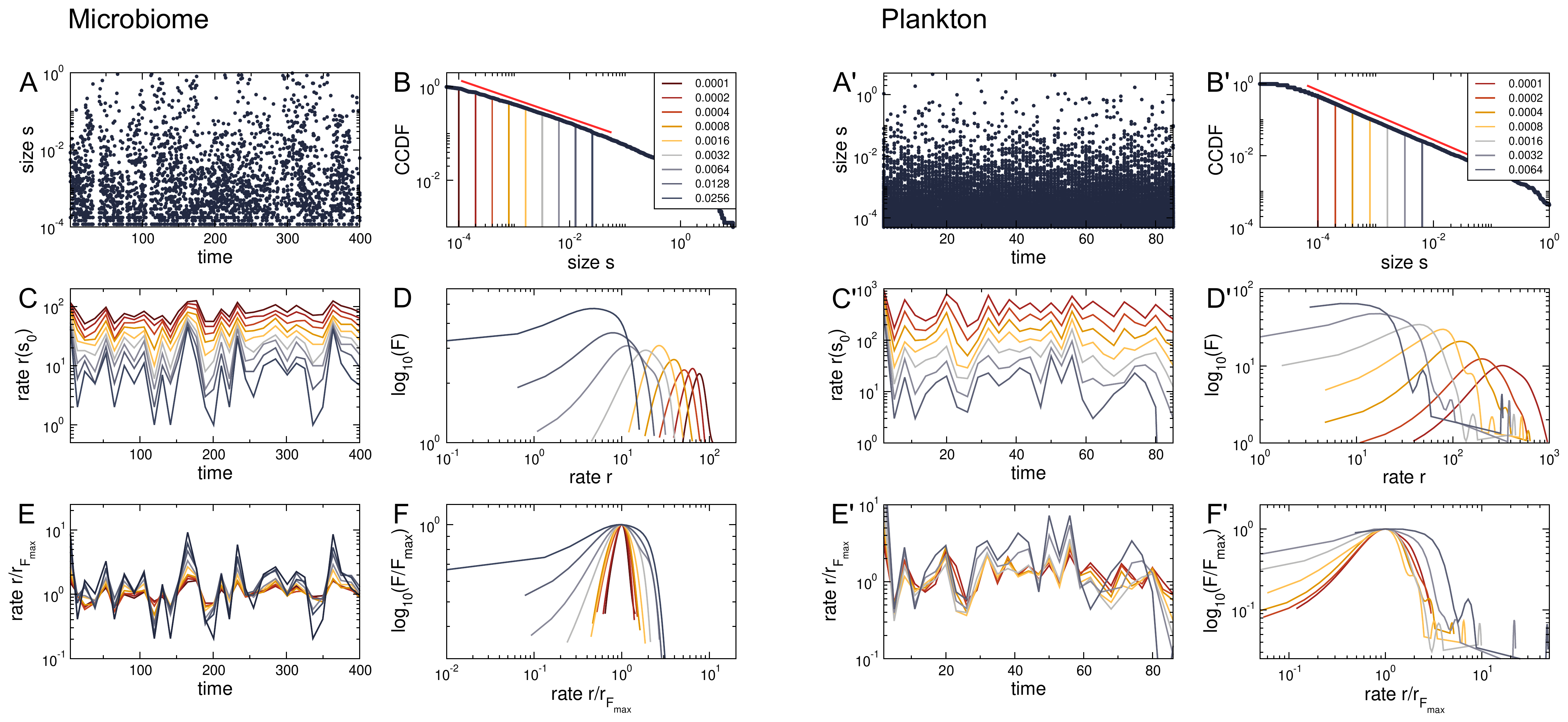}
\caption{{\bf Microbiome and Plankton Bacteria avalanching dynamics are stationary.} \emph{Left side panels: Microbiome M3} (A) Time series of consecutive avalanches sizes $s \ge 0.0001$. (B) Cumulative density distribution from data in (A). Size thresholds $s_0$ (numbers) used in C, D, E and F indicated in color. (C) Time series of avalanches rate $r(s_0)$ for sizes $s \ge s_0$ (10 day window). (D) Rate histograms for the 9 different $s_0$. (E) Normalized rate($s_0$) time series collapse into a unique series. (F) Collapsed histograms corresponding to data in (E). \emph{Right side panels: Plankton Bacteria}. Panels are formatted as in left-side panels except $s \ge s_0$ (5 day window) and the rate histograms in panel (D) are for 9 different $s_0$.}
\label{fig:fig4}
\end{figure*}
%%%%% %%%%%%%%%%%%%%%%%%%%%%%%%%%%%%%%%%%%%%%%%%%%% 

In summary, the analysis of avalanches demonstrates that the dynamics of microbial systems proceeds as a series of self-affine abundance bursts obeying specific scaling relations. Additionally, the scaling exponents obtained suggest that a critical branching process is a good candidate as the mechanism underlying microbial dynamics.

\emph{Stationarity of avalanches PDF:} It is known that a power law distribution of avalanches, as shown above, can be (hypothetically) caused by a trivial mixing of random processes with different size dependent rates.
Conversely, If microbial systems are indeed poised at criticality, then the power-law distribution of event sizes ($S$) is stationary. We test for stationarity, starting with the microbiome as in the previous section, considering first (in Fig.~\ref{fig:fig4}A) the time series of consecutive events of size $s \ge s_0$ (where $s_0 = 0.0001$). Fig.~\ref{fig:fig4}B depicts the cumulative density distribution of s. In Fig.~\ref{fig:fig4}C and D we show that, within the range where the power law scaling is valid, the rate of events, defined as the frequency of events of a certain size as a function of time, is scale invariant. This is consistent with the results in Fig.~\ref{fig:fig4}E and F showing that rate time series and rate distributions collapse after appropriately rescaling the data. Similar results hold for plankton data, as shown in the right panels of Fig.~\ref{fig:fig4}.

Two main points can be highlighted from these results: first, the stationarity of the distribution of avalanches of a wide range of sizes is in good agreement with predictions from criticality, and second, we have demonstrated that the observed event statistics are not the result of trivial mixing of random processes with different size dependent rates.

\section{Discussion}

In summary, we have analyzed the dynamics of microbial ensembles at three levels: density, correlations and avalanche dynamics. We found that the fluctuations in the density of both datasets exhibit a power law distribution over four orders of magnitude, showing in addition an anomalous scaling of the variance vs.\ mean, corresponding to an excess of variance for increasing mean abundance of the populations. This type of scaling (described usually as obeying the Taylor law \cite{eisler}) implies a higher variability than expected for a Poisson process (independent events), and is usually encountered in collective processes obeying the dynamics of critical phenomena.

We found that the correlations between the activity of the different members of the microbial communities also follow universal scaling laws very similar to those in critical systems. Specifically, the eigenvalues of the abundance correlation matrix follows finite-size scaling, implying that despite the high variability of the overall process, its correlations are scale-free, with no microbial species being more relevant or dominant than others. Furthermore, our data-driven analysis of the eigenvalues spectra predicts a hierarchical structure that mirrors very closely the taxonomic hierarchy. This is equivalent to saying that the uncovered correlations are the present result of biological evolution, selecting the microbial interactions which determine the present scale-free hierarchical biological organization in Genus, Family, etc.

An additional expression of the microbial scale-free organization is the presence of highly irregular abundance dynamics observed as large bursts of activity, also characterized by robust power law distributions with well known universal exponents. Furthermore, we demonstrate that the avalanches process is stationary, implying that the large irregularity is (almost paradoxically) very regular and constant.

Understanding the collective dynamics of microbial communities is in itself very challenging, not only because the type of data analyzed here just recently became available, but also due to the characteristic high variability they exhibit. The main novelty of the present report lies in the fact that it suggests a mechanism which dictates that high variability and robustness can be seen together in large ensembles of cooperating individuals. The three aspects described here (density fluctuations, correlations and avalanching) correspond very closely with hallmark properties of critical dynamics, suggesting that additional aspects of these correspondence deserve further study.

\emph{Acknowledgments:} Work conducted under the auspice of the Jagiellonian University-UNSAM Cooperation Agreement. Partially supported by the NIH BRAIN Initiative Grant No. 1U19NS107464-01, by CONICET (Argentina) and Escuela de Ciencia y Tecnolog\'ia, UNSAM, by UNLP (Argentina) and by the Foundation for Polish Science (FNP) project TEAMNET ``Bio-inspired Artificial Neural Networks'' (POIR.04.04.00-00-14DE/18-00). The open-access publication of this article is supported in part by the program ``Excellence Initiative - Research University'' at the Jagiellonian University in Krak\'ow (Poland).

\cleardoublepage
\pagebreak
\appendix
 
\section*{Supplementary Information}
 
%%%%%%%%%%%%%%FIG5 %%%%%%%%%%%%%%%%%%%%%%%%% 
\begin{figure}[b]
\centering
\includegraphics[width=\columnwidth]{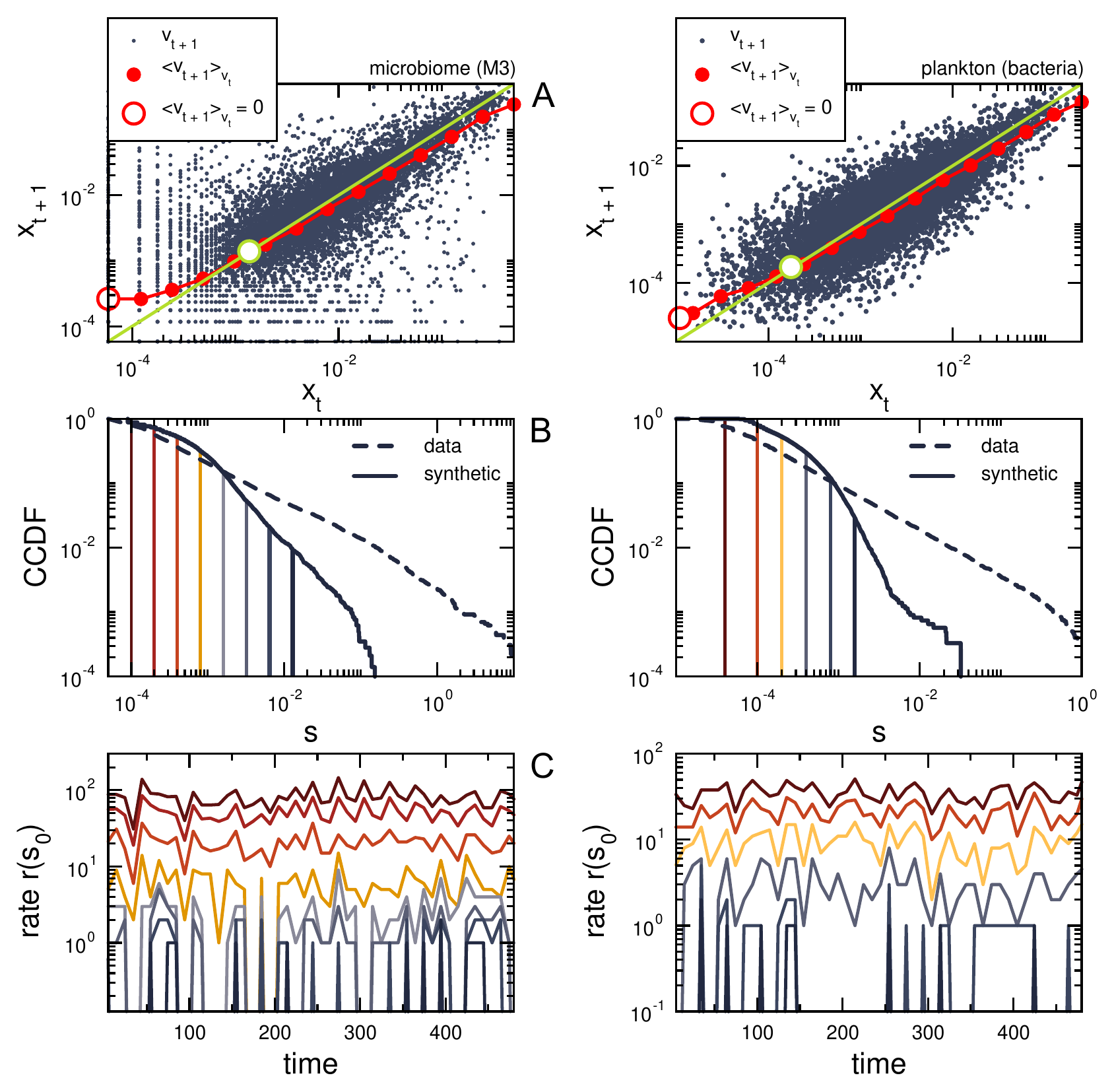}
\caption{{\bf Null model of microbial dynamics.} A) Return maps constructed using the real data from the microbiome (left) and plankton (right) by plotting consecutive normalized abundances $x(t)$ samples (dots) and its binned average $f(x(t))$ (red circles and continuous line) over-imposed. B) CCDFs from real and synthetic data and the corresponding event event sizes $s \ge s_0$ used to compute rates. C) Time series of event rate $r(s_0)$ for sizes $s \ge s_0$. }
\label{fig:fig5}
\end{figure}
%%%%%%%%%%%%%%%%%%%%%%%%%%%%%%%%%%%%%%%%%%

\emph{Additional insights from a null model:} Here we use a model which accounts for the first-order self-correlation of consecutive data points, to check how much of the statistical properties we have described can be attributed to trivial correlations. To do that, we generated synthetic time series of abundances using data from both the microbiome and the plankton, and then computed the statistics of avalanches as described in the main text.
To generate synthetic data, we built return maps using the original time series of normalized abundances. Such maps are depicted in Fig.~\ref{fig:fig5}A where all pairs of consecutive normalized abundances for all time series ($x(t)$, $x(t + 1)$) are plotted (dots). Subsequently, normalized abundances $x(t)$ were binned and its corresponding $x(t + 1)$ averaged (red circles). Similarly, to construct synthetic data, using the return map, we defined a conditional probability distribution for subsequent normalized abundances $P(x(t + 1)|x(t))$ from the data with bins of $\Delta x = 0.0005$. A special bin is used for normalized abundances after rest events (that is, abundances below the threshold $x_{th} = 0.0001$). Then, we obtained synthetic normalized abundances from a Markov process drawing from the binned distribution, according to the following procedure: (i) an initial normalized abundance $x(t)$ is chosen at random from the bin of normalized abundances following a rest event. (ii) The next normalized abundance is determined by choosing a $x(t + 1)$ at random from the corresponding bin of $P(x(t + 1)|x(t))$. (iii) Such value is then considered as the new $x(t)$ in the next iteration. (iv) If $x(t + 1) < x_{th}$, the event is considered completed. (v) If an event is finished, a rest time $t$ is randomly selected from the experimental data. (vi) After a time $t$ with $x = 0$, the procedure is restarted from (i). 
Fig.~\ref{fig:fig5}B show a comparison between CCDFs of real and synthetic data from where it is evident that the synthetic data lacks the long-range temporal correlations. Moreover, as shown in Fig.~\ref{fig:fig5}C, stationarity is lost after one decade in the synthetic data, adding weight to the idea that higher order interactions are needed to generate scale-free stationary fluctuations.

\end{document}